
\font\subtit=cmr12
\font\name=cmr8
\input harvmac
\def\LMU#1#2#3{\TITLE{LMU-TPW \number\yearltd-#1}{#2}{#3}}
\def\TITLE#1#2#3{\nopagenumbers\abstractfont\hsize=\hstitle\rightline{#1}%
\vskip 1in\centerline{\subtit #2}%
\vskip 1pt
\centerline{\subtit #3}\abstractfont\vskip .5in\pageno=0}%
\LMU{5} {ON THE QUANTIZATION OF THE CHERN$-$SIMONS FIELD THEORY}
{ON CURVED SPACE-TIMES: THE COULOMB GAUGE APPROACH}
\centerline{F. F{\name ERRARI}\foot{\name Work supported
by the Consiglio Nazionale Ricerche, P.le A. Moro 7, Roma, Italy}}\smallskip
\centerline{\it Sektion Physik der Universit\"at M\"unchen}\smallskip
\centerline{\it Theresienstr. 37, 8000 M\"unchen 2}\smallskip
\centerline{\it Fed. Rep. Germany}
\vskip 2cm
\centerline{ABSTRACT}
{\narrower
We consider here the Chern-Simons field theory with gauge group SU(N) in
the presence of a gravitational background that describes a
two-dimensional expanding ``universe".
Two special cases are treated here in detail: the spatially flat {\it
Robertson-Walker} space-time and the conformally static space-times
having a general closed and orientable Riemann surface as spatial
section. The propagator and the vertices are explicitely computed at the
lowest order in perturbation theory imposing the Coulomb gauge fixing.
}
\Date{March 1993}
\newsec { INTRODUCTION}
\vskip 1cm
In this paper we address the problem of the perturbative quantization of
the nonabelian Chern-Simons (C-S) field theory \ref\cs{S. Deser, R. Jackiw, S.
Templeton, {\it Phys. Rev. Lett.} {\bf 48} (1983), 975; {\it Ann. Phys.}
(N.Y.) {\bf 140} (1984), 372; C.R. Hagen, {\it Ann. Phys.} (N.Y.) {\bf
157} (1984), 342.}
\ref\wit{E. Witten, {\it Comm. Math. Phys.} {\bf 121} (1989), 351.}
on a curved space-time.
The space-time considered here has the form of a three dimensional
manifold $M_3$ in which the metric is induced by the following length:
\eqn\metric{ds^2=dt^2+a(t)g_{ij}(x_1,x_2)\left[dx_i\otimes
dx_j\right]}
where $g_{ij}(x_1,x_2)=h(x_1,x_2)\delta_{ij}$.
$h(x_1,x_2)$ is assumed to be a conformally flat metric on a Riemann
surface of genus $g$ $\Sigma_g$.
The resulting three dimensional
metric is static, conformally flat and with Euclidean signature.
The variable $t$ takes its values in the real line {\bf R}.
Along the $t$ axis
the metric is flat.
If the time dependent factor $a(t)$
is a constant, let say $a(t)=1$, then
we obtain the well known topological configuration
$M_3=\Sigma_g\otimes {\rm\bf R}$.
If instead $a(t)=e^{-2\rho(t)}$, $\rho(t)$ being for instance
a decreasing function, then the above metric
describes an expanding two dimensional ``universe".
\smallskip
On $M_3$ we consider the following C-S functional:
\eqn\csmatr{S_{\rm CS}={s\over 4\pi}\int_{M_3}d^3x\epsilon^{\mu\nu\rho}
{\rm Tr}\left( A_\mu\partial_\nu A_\rho+i{2\over 3}A_\mu A_\nu
A_\rho\right)}
where $A_\mu=A_\mu^a T^a$, the $T^a$ being the generators of the compact
Lie group ${\rm SU(N)}$. To evaluate the trace in eq. \csmatr,
we exploit the conventions
${\rm Tr}(T^aT^b)={1\over 2}\delta^{ab}$ and ${\rm Tr} [T^aT^bT^c]=
{1\over 4}(d^{abc}+if^{abc})$. Here $d^{abc}$ is a totally symmetric
tensor in the indices $a,b,c$ while $f^{abc}$ are the usual structure
constants of SU(N). Moreover, we have set $\mu, \nu, \rho=0,1,2$, with
$x_0=t$. $x_1$ and $x_2$ are local coordinates on the Riemann surface
$\Sigma_g$.
The indices in the spatial coordinates will be denoted by
the latin letters $i,j,k$ and so on.
Finally we remember that $s$ should be an integer in order
to preserve the gauge invariance of the theory.
A simple calculation of the trace in eq. \csmatr\ yields the following
result:
\eqn\csfunct{S_{\rm CS}={s\over 4\pi}\int_{M_3}d^3 x
\epsilon^{\mu\nu\rho}\left(A_\mu^a\partial_\nu A^a_\rho - {1\over 3}
f^{abc} A_\mu^a A_\nu^bA_\rho^c\right)}
This action is independent of the metric. The metric will be present however
in the gauge fixing action and in the Faddeev popov term.\smallskip
Despite of describing a topological field theory with no physical
degrees of freedom, the C-S functional \csfunct\ can generate non trivial
correlation functions even in the flat case,
for example in the covariant gauge
$\partial^\mu A^a_\mu=0$, $\mu=0,1,2$
\ref\guadper{E. Guadagnini,
M. Martellini, M. Mintchev, {\it Phys. Lett.} {\bf 227B} (1989), 111.}
\ref\gmm{E. Guadagnini, M. Martellini,
M. Mintchev, {\it Nucl. Phys.} {\bf B330} (1990), 575.},
\ref\alr{L. Alvarez-Gaume, J. M. F. Labastida,
A. V. Ramallo, {\it Nucl. Phys.} {\bf B334} (1990), 103.}.
Moreover, it was shown in the above references that,
computing the amplitudes of $n$ Wilson loops at any order in
perturbation theory, one can extract informations on the
HOMFLY polynomials \ref\homfly{P. Freyd, D. Yetter, J. Hoste, W.B.R.
Lickorish, K. Millet, {\it Bull. Am. Math. Soc.} {\bf 12} (1985), 239;
J.H. Przytycki, P. Traczyk, {\it Kobe Jour. Math.} {\bf 4} (1987), 115;
W.B.R. Lickorish, K. millet, {\it Topology} {\bf 26} (1987), 107.}
from the C-S field theory.\smallskip
Unfortunately, it is not easy to perform analogous computations in the
case of a non-flat space-time. For instance, in the covariant gauge
$\partial^\mu A^a_\mu=0$, the correlation functions of the C-S field
theory defined on a curved metric background are in general not known.
In the temporal gauge $A_0=0$, instead, one has to solve explicitely the
Gauss constraint that fixes the residual gauge invariance
\wit, \ref\djt{G.V. Dunne, R. Jackiw, C.K. Trugenberger, {\it Ann. of
Phys.} {\bf 194} (1989), 197.}.
In the presence of Wilson loops this becomes a difficult task due to the
zero modes \ref\ilr{J.M. Isidro, J.M.F. Labastida, A.V.
Ramallo, Preprint US-FT-9-92.} \ref\bn{M. Bos, V.P. Nair, {\it Int.
Jour. Mod. Phys.} {\bf A5} (1990), 959.}
and, in fact, we believe that this problem has not been solved until
now.
Finally, the light-cone gauge of ref. \ref\fk{J. Fr\"ohlich, C. King,
{\it Comm. Math. Phys.} {\bf 126} (1989), 167.} is not
compatible with the transition functions of
a manifold like that described by the metric
\metric.\smallskip
For these reasons, we propose here a perturbative approach to the C-S
field theory quantized in the Coulomb gauge.
As a noncovariant gauge, the Coulomb gauge has not yet been investigated
in connection with the C-S theory, as for example the temporal or
light cone gauges, which have already been widely studied
\ref\noncovgau{T. Gajsosik, W. Kummer, Preprint TUW-92-19; A.N.
Kapustin, P.I. Pronin, {\it Phys. Lett.} {\bf  B274} (1992), 363; G.
Leibbrandt, C.P. Martin, {\it Nucl. Phys.} {\bf B377} (1992), 593; S.
Emery, O. Piguet, {\it Helv. Phys. Acta} {\bf 64} (1991), 1256.}.
In this gauge, however, the quantization in a curved space-time
endowed with the metric \metric\ will be
drastically simplified as we will see.\smallskip
The material of this paper is divided as follows.
In Section 2 we quantize the C-S field theory in a
spatially flat {\it
Robertson-Walker} space-time. The explicit expressions of the propagators
of the gauge fields and of the ghosts together with
those of the vertices is given in the Coulomb gauge.
In Section 3 these results are generalized to the case of a three
dimensional manifold
with conformal metric. The basic spatial section of the manifold is a
closed and orientable Riemann surface of genus $g$.
Finally, we discuss in the Conclusions some of the possible applications
of the perturbative approach presented here.
\smallskip
\vskip 1cm
\newsec{THE CHERN-SIMONS FIELD THEORY IN THE COULOMB GAUGE: THE FLAT CASE}
\vskip 1cm
First of all we consider the quadratic part of the action \csfunct\ in order
to compute the propagators of the vector potentials $A_\mu$. We remember that
throughout this Section we consider only the flat case, i.e.
$g_{ij}(x_1,x_2)=\delta_{ij}$ in eq.\metric.
Firstly, we have to fix the gauge invariance of the C-S theory.
A gauge transformation
\eqn\potgt{A'_\mu=UA_\mu U^{-1}+\partial_\mu U U^{-1}}
is generated
by the elements of SU(N):
\eqn\groupel{U(x_1,x_2,t)={\rm exp}\left[\omega^a(z,\bar z,t)T^a\right]}
where the $\omega^a(x_1,x_2,t)$ represent functions on $M_3$.
A convenient gauge fixing is the Coulomb
gauge
\eqn\cugauge{\partial^i A_i^a=0}
The coordinates $x_1$ and $x_2$ are in this case global coordinates on
the two dimensional plane ${\rm\bf R}^2$.
In order to quantize the theory, we have to insert as usual a gauge-fixing
term in the action \csfunct\ and the Faddeev popov ghosts:
\eqn\sq{S_{\rm q}=S_{\rm CS}+S_{\rm gf}+S_{\rm fp}}
where $S_{\rm CS}$ was already defined above and
\eqn\sgf{S_{\rm gf}={s\over 8\pi}
\int_{M_3}d^3x\sqrt{g}{1\over \lambda}(\partial^i
A_i^a )^2}
\eqn\sfp{S_{\rm fp}=\int_{M_3}d^3x\sqrt{g}\bar
c^a\left[\partial^iD^{ab}_i(A)\right]c^b}
In eq. \sgf\
$\lambda$ represents an arbitrary real parameter.
Moreover, $D^{ab}_\mu(A)$ in eq. \sfp\ denotes the usual covariant
derivative:
\eqn\covdev{D_\mu^{ab}(A^c)=\partial_\mu\delta^{ab}-f^{abc}A^c_\mu}
and $\sqrt{g}=|det(g_{\mu\nu})|^{1\over 2}$. In the flat case we are
treating,
$\sqrt{g}=a(t)$.
Finally, $\bar c^a(z,\bar z,t)$ and $c^b(z,\bar z,t)$ represent the conjugate
ghost fields. As we see from eqs. \sgf\ and \sfp, the  gauge fixing
and the Faddeev-Popov Lagrangians depend on the metric. In our flat
case, this means that there is a dependence on the function $a(t)$
defined in eq. \metric.
Of course, eq. \covdev\ is only valid in a local sense, since in the
case of a general manifold one needs to introduce also the Christoffel
symbols in order to made it invariant under diffeomorphisms.\smallskip
The expression of the propagator in the Coulomb gauge can be obtained
solving the following equation:
\eqn\diffeq{{s\over 8\pi}
\left[\epsilon^{\mu\nu\rho}\partial_\nu+\left(\lambda
a_{x_0}\right)^{-1}
\left(\partial^\mu
-\eta^\mu(\eta\cdot\partial)\right)\left(\partial^\rho
-\eta^\rho(\eta\cdot\partial)\right)
\right]
G_{\rho\kappa}^{ab}(x-y)=\delta^{ab}\delta^\mu_\kappa\delta^{(3)}(x-y)}
where $\int d^3x\delta(x)f(x)=f(0)$.
Again, this equation is valid only in the flat case.
In eq. \diffeq\ $\eta_\mu$ denotes the vector
$\eta_\mu=(0,0,1)$ in the three dimensional space ${\rm\bf R}^3$.
Solving eq. \diffeq\ gives the following components of the
propagator:
\eqn\gbg{G_{ij}^{ab}(x-y)=8\pi\lambda a(x_0)\delta^{ab}
\delta(x_0-y_0)\partial_i\partial_j \Delta_2(\overline x-\overline y)}
\eqn\gbzero{G_{0i}^{ab}(x-y)=+{8\pi\over
s}\delta^{ab}
\left[
\delta(x_0-y_0)\epsilon_{ij} \partial^j\Delta_1(\overline
x-\overline y)-\lambda\partial_{x_0}
\left(
\delta(x_0-y_0)
a(x_0)
\right)
\partial_i
\Delta_2(\overline x-\overline y)
\right]
}
\eqn\gzerob{G_{i0}^{ab}(x-y)=
{8\pi\over s}
\delta^{ab}
\left[
\delta(x_0-y_0)
\epsilon_{ij}
\partial^j
\Delta_1(\overline x-\overline y)
-\lambda
a(x_0)
\partial_{x_0}
\delta(x_0-y_0)
\partial_i
\Delta_2(\overline x-\overline y)\right]}
\eqn\gzerozero{G_{00}^{ab}(x-y)=-{8\pi\lambda\over
s}\delta^{ab}\partial_{x_0} \left[{a(x_0)\partial_{x_0}\delta(x_0-y_0)
 }\right]\Delta_2(\overline x-\overline y)
}
Here $\epsilon_{\alpha\beta}$ is the two dimensional $\epsilon$ tensor
with $\epsilon_{12}=1$ and $\overline x=(x_1,x_2)$.
$\Delta_1(x-y)$ is the usual propagator of the massless scalar fields:
\eqn\gfmsf{\Delta_1(\overline x-\overline y)=
\int {d^2\overline k\over
(2\pi)^2}{e^{i\overline k\cdot(\overline x-\overline y)}\over \overline
k^2}}
and
\eqn\gfmbh{\Delta_2(\overline x-\overline y)=
\int {d^2\overline k\over
(2\pi)^2}{e^{i\overline k\cdot(\overline x-\overline y)}\over \overline
k^4}}
where $\overline k=(k_1,k_2)$.
It is important to stress here that the propagator $G_{ij}^{ab}(x-y)$
is longitudinal in both indices $i$ and $j$. This descends from eq.
\diffeq\ after
setting $\mu=0$ and $\kappa=j$:
\eqn\longit{\epsilon^{0lj}\partial_lG_{ij}^{ab}(x-y)=0}
\smallskip
The propagator in the Coulomb gauge can be obtained performing the limit
$\lambda=0$. In this limit, it is easy to see that all the component of the
propagator vanish but the transverse components of
$\tilde G_{0i}^{ab}(x-y)$
and $\tilde G_{i0}^{ab}(x-y)$.
Therefore, in the Coulomb gauge, the propagators are given by:
\eqn\propfina{G^{ab}_{0i}(x-
y)={8\pi\over s}\delta^{ab}\delta(x_0-y_0)\epsilon_{ij}
\partial^j \Delta_1(\overline x-\overline y)}
\eqn\propfinb{G^{ab}_{0i}(x-y)=G^{ab}_{i0}(x-y)}
In the same way one can compute the propagator of the ghost fields
$G^{ab}_{gh}(x,y;t,t')$, which reads:
\eqn\ggh{G^{ab}_{gh}(x-y)=\delta^{ab}\Delta_1(\overline x-
\overline y)\delta(x_0-y_0)}
Now we are ready to compute the vertices of the C-S field theory in the Coulomb
gauge.
As we see from the expressions of the propagators \propfina\ and
\propfinb, the
factor $8\pi\over s$ can be viewed as a coupling constant (see for example
\guadper\ on this point).
With the settings $x=(\overline x,x_0)$, $y=(\overline y,y_0)$, $z=(
\overline z,z_0)$ and $w=(\overline w,v)$ and at the first order in
perturbation theory,
the amplitude between three gauge
fields $A_\mu^a(x)$, $A_\nu^b(y)$ and $A_\rho^c(z)$ is:
\eqn\vertdef{V_{\mu\nu\rho}^{abc}(x,y,z)=-{s\over 24\pi}\int dw^3 f^{def}
\epsilon^{\sigma\tau\eta}<A_\mu^a(x)A_\nu^b(y)A_\rho^c(z)A_\sigma^d(w)
A_\tau^e(w)A_\eta^f(w)>}
It is easy to see that in the Coulomb gauge only those components of the
vertex
survive, for which two of the indices $\mu$, $\nu$ and $\rho$ are equal to
zero, while the third index is a spatial index $i$:
\eqn\vertspat{V_{i00}^{abc}(x,y,z)=-{s\over 6\pi}\int d^{\overline w}dv
f^{def}
\epsilon^{0jk}G_{i0}^{ad}(\overline x,\overline w;x_0,v)
G_{0j}^{be}(\overline y,\overline w;y_0,v)
G_{0k}^{cf}(\overline z,\overline w;z_0,v)}
In principle we had to use the Levi Civita tensor
$[\epsilon]^{\mu\nu\rho}= a(x_0)\epsilon^{\mu\nu\rho}$
in eq. \vertspat. However, the dependence on the metric $a(x_0)$
cancels against the determinant of the metric present in the integration
measure $d^3w$, so that in the final form of the vertex the metric
disappears.
Let us also notice that, since each of the external propagator is proportional
to the factor $8\pi/s$, the effective coupling constant in front of the
vertex \vertspat\ is going as $1\over s^2$.
Finally, we have to compute the vertex of the interaction between ghosts and
gauge fields.
A straightforward calculation yields:
\eqn\ghostvert{V_{0\enskip {\rm gh}}^{abc}(x,y,z)=\int d\overline w^2dv a(v)
f^{def}G_{\rm gh}^{ad}(x-w)\partial^{i}_{(w)}G_{\rm
gh}^{be}(x-w)G_{0i}^{cf}(z-w)}
All the other components of this vertex vanish.
As it is possible to see after summing over the index $i$ using the
expression of the metric \metric\ in the flat case,
the ghost vertex does not depend on $a(t)$.
\vskip 1cm
\newsec{THE CHERN-SIMONS FIELD THEORY IN THE COULOMB GAUGE:
THE CASE OF CURVED SPACE-TIMES}
\vskip 1cm
Let us consider the most general manifold $M_3$ in which the metric is given
by eq. \metric.
The spatial section $\Sigma_g$ is a Riemann surface of genus $g$.
Covering $\Sigma_g$ with a system of open sets $\{U_i\}$, we define the
local coordinates $z^{(i)},\bar z^{(i)}:U_i\rightarrow{\rm\bf C}$ as
follows:
\eqn\locoord{\matrix{z^{(i)}=x_1^{(i)}+ix_2^{(i)}\cr
\bar z^{(i)}=x_1^{(i)}-ix_2^{(i)}\cr}}
{}From now on, we will drop out the subscript $(i)$.
The only nonvanishing components of the metric \metric\ are in these
coordinates:
\eqn\compmetric{
g_{00}=1\qquad\qquad g_{z\bar z}(z,\bar z,t)=g_{z\bar z}(z,\bar z,t)=a(t)
h(z,\bar z)\qquad\qquad g^{z\bar z}g_{z\bar z}=1}
The gauge fixed
Chern-Simons action becomes in complex coordinates $S_{\rm CS}=S_{\rm
free}+S_{\rm int}$, where:
$$S_{\rm free}=
\int_{M_3}
d^2zdt
\left[
2i
\left(
A_0^a\partial_{\bar z}A_z^a+
A_z^a\partial_0A_{\bar z}^a+
A_{\bar z}^a\partial_zA_0^a-
{\rm c.c.}
\right)
\right.$$
\eqn\csfree{
\left.
+\left(
a(t)\lambda
\right)^{-1}
g^{\bar z z}
(\partial_zA_{\bar z}^a+
\partial_{\bar z}A_z^a)^2
+2\bar c^a\partial_z\partial_{\bar z}c^a\right]}
\eqn\csint{S_{\rm
int}=\int_{M_3}d^2zdt\left[\epsilon^{\mu\nu\rho}f^{abc}
A_\mu^aA_\nu^bA_\rho^c -f^{abc}\bar c^a\left(A_z^b\partial_{\bar
z}+A_{\bar z}^b\partial_z\right)c^c\right]}
and $d^2z={1\over 2i}dz\wedge d\bar z$.
The factor $2i$ in eq. \csfree\ comes from the form of the
$\epsilon^{\mu\nu\rho}$ tensor in complex coordinates.
In fact, the Levi-Civita tensor $[\epsilon]^{\mu\nu\rho}=g^{-{1\over 2}}
\epsilon^{\mu\nu\rho}$ becomes in these coordinates:
\eqn\levicivita{[\epsilon]^{0z\bar z}=-2ig^{z\bar z} a^{-1}(t)}
All the other components can be obtained from eq. \levicivita\
permuting the indices $0$, $z$ and $\bar z$ and changing the sign
according to the order of the permutation.
In this Section it will be useful to denote a sum over the complex
indices with the first letters
of the Greek alphabet $\alpha,\beta,\gamma$ and so on.
For example, the gauge condition \cugauge\ becomes now $\partial^\alpha
A_\alpha^a=0$. Using the metric \metric\ to rise and lower the indices,
this equation reads:
\eqn\complgauge{\partial_zA_{\bar z}^a+\partial_{\bar z}A_z^a=0}
Eq. \complgauge\ does not contain the metric explicitely. This means that the
Coulomb gauge condition is compatible with the transition functions at
the intersections $U_i\cap U_j$ of two open sets $U_i$ and $U_j$ of the
covering of $\Sigma_g$. Therefore eq. \complgauge\ is globally valid on
$M_3$.\smallskip
The gauge fields $(A_z^a,A_{\bar z}^a,A_0^a)$ are connections on the
trivial principal bundle $$P(M_3,{\rm SU(N)})=M_3\otimes{\rm SU(N)}$$
This bundle is trivial due to the fact that SU(N) is a simply connected
Lie group.
One can show as in the flat case that the Coulomb gauge \complgauge\ is
a good gauge fixing without Gribov ambiguities
\ref\gribov{V. N. Gribov, {\it Nucl. Phys.} {\bf B139} (1978),1;
P. van Baal, {\it Nucl. Phys.} {\bf 369} (1992), 259.}, at least in the
perturbative approach.
A proof can be given as in \ref\iz{C. Itzykson,
J-B. Zuber, Quantum Field Theory, McGraw-Hill 1980, p. 576.},
considering a
gauge transformation \groupel\ in which $\omega^a(z,\bar z,t)$ is taken
to be an infinitesimal small function.
The new fields, after this gauge transformation, are of the form
$A_\mu^{a\prime}=A_\mu^a+(D_\mu\omega)^a$.
The requirement that also $A_\mu^{a\prime}$ satisfies the gauge condition
\complgauge\ yields the following equation:
\eqn\cucons{\partial_z D^{ab}_{\bar z}(A)\omega^b(z,\bar z,t)+
\partial_{\bar z} D^{ab}_z(A)\omega^b(z,\bar z,t)=0}
If we suppose that $A_\mu^a$ is only a small perturbation around a
classical solution, then we can
expand the functions $\omega^a(z,\bar z,t)$ in powers of the effective
coupling constant $1\over s$
$$\omega^a(z,\bar z,t)=\sum\limits_{n=0}^\infty
\left({1\over s}\right)^n\omega^a_{(n)}(z,\bar z,t)$$
Solving eq. \cucons\ in terms of the $\omega_{(n)}(z,\bar z,t)$,
we get as a general solution on $M_3$:
$$\omega_{(n)}^a(z,\bar z,t)=\omega_{(n)}(t)$$
This is due to the fact that the spatial section of $M_3$ is a compact
Riemann surface.
Still we have to set the boundary conditions of the fields $A_z^a(z,\bar
z,t)$.
Requiring that the fields $A_z^a$ vanish at infinity, i.e. for large
values of $|t|$, it is easy to see that the functions $\omega_{(n)}(t)$
should be constant. Therefore, the residual gauge invariance after the
Coulomb gauge fixing amounts to the constant elements of the group
SU(N) and it is not difficult
to integrate it out in the path
integral.\smallskip
Having proven that the Coulomb gauge fixing is valid also in the case of
the manifold $M_3$ with metric \metric,
we are ready to compute the propagators of the gauge
fields.
Let us put
$$G^{ab}_{\mu\nu}(z,w;t,t')=<A^a_\mu(z,\bar z,t)A^b_\nu(w,\bar w,t)>$$
where now $\mu,\nu=0,z,\bar z$.
Then the equations satisfied by the propagators of the $A-$fields become:
\eqn\propI{-4i\partial_z
G^{ab}_{\bar z0}(z,w;t,t')
+4_i\partial_{\bar
z}
G^{ab}_{z0}(z,w;t,t')
={8\pi\over s}\delta^{ab}\delta^{(2)}_{z\bar z}(z,w)\delta(t-t')}
$$-4i\partial_{\bar
z}
G^{ab}_{0w}(z,w;t,t')
+4i\partial_0
G^{ab}_{\bar z w}(z,w;t,t')
-2{a^{-1}(t)\over \lambda}\partial_{\bar
z}\left[g^{z\bar z}\partial_{\bar z}
G^{ab}_{zw}(z,w;t,t')
+\right.$$
\eqn\propII{
\left.+
g^{z\bar z}\partial_z
G^{ab}_{\bar z w}(z,w;t,t')
\right]={8\pi\over s}\delta^{ab}\delta^{(2)}_{\bar z w}
(z,w)\delta(t-t')}
Another equation can be obtained from \propII\
permuting the indices $z$ and $\bar z$ and substituting the index $w$
with $\bar w$.
There are still other relations relating the various components of the
propagators together:
\eqn\propIII{\partial_z
G^{ab}_{\bar z \alpha}(z,w;t,t')
-
\partial_{\bar z}G^{ab}_{z\alpha}(z,w;t,t')
=0}
$$-4i\partial_\alpha
G^{ab}_{00}(z,w;t,t')
+4i\partial_0
G^{ab}_{\alpha0}(z,w;t,t')
-$$
\eqn\propIV{-{a^{-1}(t)\over
\lambda}\partial_\alpha\left[ g^{z\bar z}\partial_{\bar
z}
G^{ab}_{z0}(z,w;t,t')
+g^{z\bar z}\partial_z
G^{ab}_{\bar z 0}(z,w;t,t')
\right]=0}
where $\alpha=w,\bar w$. Eq. \propIII\ implies that the propagators
$G_{z\bar z}^{ab}(z,w;t,t')$ and $G_{\bar z z}^{ab}(z,w;t,t')$
do not have transverse components. This equation is the equivalent in
complex coordinates of
the relation \longit.
Finally we have:
$$-4i\partial_{\bar z}
G^{ab}_{0\bar w}(z,w;t,t')
+4i\partial_0
G^{ab}_{\bar z\bar w}(z,w;t,t')
-$$
\eqn\propV{-{a^{-1}(t)\over \lambda}\partial_{\bar z}\left[g^{z\bar
z}\partial_{\bar z}
G^{ab}_{z\bar w}(z,w;t,t')
+g^{z\bar
z}\partial_z
G^{ab}_{\bar z\bar w}(z,w;t,t')
\right]=0}
Again it is possible to get another independent relation from eq.
\propV\ interchanging the two indices $z$ and $\bar z$ and substituting
$\bar w$ with $w$.
Eqs. \propI-\propV\ are equivalent to the system \diffeq\ given in the
flat case.
It is very difficult to solve these equations when the metric is the
general metric given in eq. \metric.
Moreover, we should remember that due to  a theorem stating that the
total charge on a Riemann surface (like in any other two dimensional
compact manifold) is always zero, an isolated $\delta$ function
$\delta^{(2)}(z,w)$ is not allowed. Therefore, in the right hand sides
of eqs. \propI-\propII\ there must be also zero modes.
The form of these zero modes will be uniquely determined below.
Despite of all these difficulties,
in the Coulomb gauge, i.e. in the limit in which
$\lambda\rightarrow 0$, there are drastic simplifications,
so that the above equations are reduced to the following
two relations:
\eqn\simpleone{\partial_z
G^{ab}_{\bar z0}(z,w;t,t')
-\partial_{\bar
z}
G^{ab}_{z0}(z,w;t,t')
={4\pi i\over
s}\delta^{ab}\delta^{(2)}(z-w)\delta(t-t')+{\rm zero}\enskip{\rm modes}}
\eqn\simpletwo{\partial_{\bar
z}
G^{ab}_{z0}(z,w;t,t')
+\partial_z
G^{ab}_{\bar z 0}(z,w;t,t')
=0}
These equations describe exactly the main requirement of the Coulomb
gauge, i.e. the fact that only the transverse fields in the two
dimensional spatial section $\Sigma_g$ of $M_3$ propagate.
The transverse fields in complex coordinates are in fact described by
the following condition: $A_z^a=\overline{(A_z^a)}=-A_{\bar z}^a$.
The solution of eqs. \simpleone\ and \simpletwo\ is provided by the
following Green functions:
\eqn\azaz{<A_z^a(z,t)A_0^b(w.t')>={2\pi i\over s}
\delta^{ab}\partial_zK(z,w)\delta(t-t')}
\eqn\azbaz{<A_{\bar
z}^a(z,t)A_0^b(w,t')>=-{2\pi i\over s}
\delta^{ab}\partial_zK(z,w)\delta(t-t')}
where $K(z,w)$ is the usual propagator of the scalar fields on a Riemann
surface (see ref.
\ref\vv{E. Verlinde, H. Verlinde, {\it Nucl. Phys.} {\bf B288}
(1987), 357.} for more details):
\eqn\dzdzbk{K(z,w)=\delta_{z\bar z}^{(2)}(z,w)+{g_{z\bar z}\over
\int_{\Sigma_g} d^2ug_{u\bar u}}}
\eqn\dzdwbk{
\partial_z\partial_{\bar w}K(w,z)=-\delta^{(2)}_{z\bar w}(z,w)
+\bar\omega_i(\bar
z)\left[{\rm Im}\enskip \Omega\right]^{-1}_{ij}\omega_j(w)}
\eqn\ik{\int_{\Sigma_g}d^2zg_{z\bar z}K(z,w)=0}
In eq. \dzdwbk\ the $\omega_i(z)dz$, $i=1,\ldots,g$, denote the usual
holomorphic differentials and $\Omega_{ij}$ represents the period
matrix.
It is important to stress here that $K(z,w)$ is a singlevalued function
on $\Sigma_g$.
Using the propagators \azaz\ and \azbaz\ it is easy to see that eq.
\simpletwo\ is trivially satisfied.
Therefore, the Coulomb gauge requirement \complgauge\
is fulfilled and the above defined propagators describe exactly the transverse
components of the gauge fields.
Still there is an ambiguity in the solutions \azaz\ and \azbaz\ due to
the zero mode sector of the fields $A_z^a$ and $A_{\bar z}^a$.
In order to remove this ambiguity, we have to require that the above
propagators are singlevalued along the nontrivial homology cycles of the
Riemann surface. Otherwise, the propagators are not well defined on
$M_3$, but in one of its coverings.
Therefore, the propagators should obey the following relations:
\eqn\zeroholonomy{\oint_{\gamma}dz<A_z^a(z,t)A_0^b(w,t')>=
\oint_{\bar \gamma}d\bar z<A_{\bar z}^a(z,t)A_0^b(w,t')>=0}
along any nontrivial homology cycles $\gamma$.
Due to the properties of singlevaluedness of the Green function $K(z,w)$,
eq. \zeroholonomy\ is trivially satisfied by the propagators given in
eqs. \azaz\ and \azbaz.
In this way these two propagators are well defined and also the freedom
in the zero mode sector is removed.
Now we insert their expressions in eq. \simpleone\ in order to get
the exact form of the zero mode terms appearing
in the right hand side of this
equation:
\eqn\simplezm{\partial_z
G_{\bar z0}^{ab}(z,w;t,t')
-\partial_{\bar
z}
G_{z0}^{ab}(z,w;t,t')
={4\pi i\over
s}\delta^{ab}\delta^{(2)}(z,w)\delta(t-t')+ {4\pi is}{g_{z\bar z}
\over \int_{\Sigma_g} d^2ug_{u\bar u}}\delta(t-t')}
The fact that the propagators in the Coulomb gauge must obey eq.
\zeroholonomy\ can be understood also decomposing the fields by means of
the Hodge decomposition of the gauge fields in a coexact, exact and
harmonic part:
\eqn\azdec{A_z^a=i\partial_z\varphi^a+\partial_z\rho^a+A_z^{\rm har}}
\eqn\azbdec{A_{\bar z}^a=i\partial_{\bar z}\varphi^a+\partial_{\bar z}
\rho^a+A_{\bar z}^{\rm har}}
$\phi^a$ and $\rho^a$ represent two real scalar fields.
The above decomposition is allowed since the gauge invariance has been
completely fixed by the choice of the Coulomb gauge, at least in the
perturbative approach, and the $G-$bundle $P(M_3,{\rm SU(N)})$ is
trivial as we previously remarked.
In the Coulomb gauge, the only components of the gauge fields which are
allowed to propagate are the coexact differentials, i.e. the
$1-$forms obtained differentiating the scalar fields $\varphi^a$ in eqs.
\azdec\ and \azbdec. Therefore, the requirement \zeroholonomy\
is a pure consequence of the fact that
the coexact forms have vanishing holonomies around the nontrivial
homology cycles.\smallskip
Let us notice that the zero mode term appearing in the right hand side
of eq. \simplezm\ is totally irrelevant.
To eliminate it it is sufficient to introduce new gauge fields, let
say $\tilde A_z$, $\tilde A_{\bar z}$, differing from the old ones
by the fact that they are normalized to zero at a point $(0,0)$ of the
Riemann surface\foot{On $M_3$ this implies that the new fields are
normalized to zero along the whole line of the time. This is possible to
do since the three dimensional manifold
is flat in the time direction.}:
\eqn\newaz{\tilde A_z^a(z,\bar z,t)=A_z^a(z,\bar z,t)-A_z^a(0,0,t)}
\eqn\newazb{\tilde A_{\bar z}^a(z,\bar z,t)=
A_{\bar z}^a(z,\bar z,t)-A_{\bar z}^a(0,0,t)}
Using the above new fields it is easy to check that the second term
in the right hand
side of eq. \simplezm, which is a zero mode contribution, cancels out.
Finally, we notice that in the flat case discussed in Section 2, the
propagators \azaz\ and \azbaz\ are in agreement with the propagators given in
eqs. \propfina\ and \propfinb.\smallskip
We finish this Section providing the explicit form of the other correlation
functions of Chern-Simons field theory.
The propagator of the ghost fields becomes:
\eqn\gghrs{G_{\rm gh}^{ab}(z,w;t,t')=\delta^{ab}K(z,w)\delta(t-t')}
The vertex coming from the cubic interaction between the gauge fields
reads instead:
$$V_{z_100}^{abc}(z_1,z_2,z_3;t,t',t'')=
{2\pi^2 is\over
3}\int_{\Sigma_g}
d^2zf^{abc}\partial_{z_1}K(z_1,z)\left[\partial_zK(z_2,z)\partial_{\bar
z}K(z_3,z)-\right.$$
\eqn\cmvertex{\left.\partial_{\bar
z}K(z_2,z)\partial_zK(z_3,z)\right]\delta(t-t'')\delta(t'-t'')}
The simple integration in the variable $t$ has been already carried out
in the above expression of the vertex.
The component
$V^{abc}_{\bar z_1 00}(z_1,z_2,z_3;t,t',t'')$ of the vertex can be
simply obtained replacing the derivative $\partial_{z_1}$ in eq.
\cmvertex\ with its complex conjugate.
The vertex describing the interaction between ghost and gauge fields has
only one component which is given by:
$$V_{0\enskip {\rm gh}}^{abc}(z_1,z_2,z_3;t,t',t'')={2\pi
i\over sa(t)}
\int_{M_3}d^2zf^{abc}K(z_1,z)\left[\partial_zK(z_2,z)\partial_{\bar
z}K(z_3,z)\right.$$
\eqn\cmvghzero{\left.-\partial_{\bar z}K(z_2,z)\partial_zK(z_3,z)\right]
\delta(t-t')\delta(t'-t'')}
It is easy to check that the above expressions of the
vertices \cmvertex\ and \cmvghzero\ are real as it should be.
\vskip 1cm
\newsec{CONCLUSIONS}
\vskip 1cm
Concluding, we would like to outline some of the possible applications
of the perturbative approach presented here.
On one side, the introduction of the C-S functional has been proposed
in condensed matter physics as
a possible explanation of some physical phenomena, like for example
the fractional quantum Hall effect \ref\fqhe{J. Fr\"ohlich, T. Kerler,
P.A. Marchetti, {\it Nucl. Phys.} {\bf B374} (1992), 511; J. Fr\"ohlich,
A. Zee, {\it Nucl. Phys.} {\bf B364} (1991), 517; X.G. Wen, {\it Phys.
Rev.} {\bf B40} (1989), 7387; B. Blok, X.G. Wen, {\it Nucl. Phys.} {\bf
B374} (1992), 615; R. Iengo, K. Lechner, {\it Phys. Rep.} {\bf 213}
(1992), 179; M. Bergeron, D. Eliezer, G. Semenoff, Preprint
UBCTP-92-001.}.
We expect therefore
that the presence of a background can be revealed by some new and
observable effects.
For instance, it was proven in \ref\scalars{F. Ferrari, {\it Phys.
Lett.} {\bf 277B} (1992), 423; Preprint LMU-TPW 92-13, to be published
in {\it Comm. Math.
Phys.}; Preprint LMU-TPW 92-24.} that a curved
space-time manifests itself in a two dimensional scalar field theory
through the appearance in the amplitudes of induced vertex operators
satisfying a nonabelian braid group statistics.
Unfortunately this example, unlike the C-S field theory, is unphysical,
apart from its consequences in string theory.\smallskip
In the C-S field theory we suppose that the mechanism through which
topological effects become evident, is provided by the edge states of
ref. \ref\balac{A.P. Balachandran, G. Bimonte, K.S. Gupta, A. Stern,
{\it Int. Jour. Mod. Phys.} {\bf A7} (1992), 5855; {\it ibid.} {\bf A7}
(1992), 4655; A.P. Balachandran, R.D. Sorkin, W.D. McGlinn, L.
O'Raifeartaigh, S. Sen, {\it Int. Jour. Mod. Phys} {\bf A7} (1992),
6887; A.P. Balachandran, Preprint SU-4240-506.}.
In fact, one can always choose the metric \metric\ in such a way that
the curvature of the Riemann surface $\Sigma_g$ has point-like
singularities at some points $a_1,\ldots,a_M\in \Sigma_g$
(see ref. \scalars\ for
more details).
These points play the role of punctures and, therefore, they generate
edge states with nontrivial statistics in the presence of the C-S fields.
Preliminary calculations, done for the abelian C-S theory, show indeed
that edge states of this kind are induced by the punctures
$a_1,\ldots,a_M$ as the vertex operators are induced in the two
dimensional scalar model mentioned above.\smallskip
On the other side, the nonabelian C-S field theory is currently being
studied on curved space-times due to its applications to knot theory
\wit, \ref\knot{M.F. Atiyah, in Proc. Symp. Pure Math. {\bf 48}, ed. R.
Wells, ({\it Am. Math. Soc.} (1988); A. Schwarz, {\it Lett. MAth. Phys.}
{\bf 2} (1978), 247; {\it Comm. Math. Phys.} {\bf 67} (1979), 1.}
and braid group statistics
\fk, \balac, \ref\bgs{ J. Fr\"ohlich, F. Gabbiani, P.A. Marchetti, in
BANFF NATO ASI (1989), 15; G. Felder, J. Fr\"ohlich, G. Keller, {\it
Comm. Math. Phys.} {\bf 124} (1989), 647; J. Fr\"ohlich, in Proc. of the
Gibbs Symposium, Yale University, D.G. Caldi, G.D: Mostow (eds.) (1990).}.
For instance, the perturbative approach developed here makes it possible
to derive the multiparameter link invariants of refs.
\ref\turaev{V.G. Turaev, in C. Yang and M.L. Ge (eds),
{\it Braid Groups, Knot Theory and Statistical Mechanics}, World
Scientific, Singapore, (1989).} and \ref\pcrr{P. Cotta-Ramusino, M.
Rinaldi, in Proc. of the XXth Int. Conf. on Diff. Geom. Methods in
Theor. Phys., S. Catto, A. Roche (eds.), World Scientific 1992.}
from a nonabelian C-S field theory
defined on a 3-manifold $\Sigma_g\otimes {\rm\bf R}$.
This is a very interesting application since
these link invariants were obtained until now only
within the framework of the
quasi-triangular Hopf algebras, but a field theoretic approach is still
missing.
Finally, we notice that the manifold $\Sigma_g\otimes {\rm\bf R}$
is just a subcase of the manifolds $M_3$ treated here, which can be
obtained setting $a(t)=1$ in eq. \metric.
However, as we have seen here, the dilation factor $a(t)$ does not
appear in the correlation functions of the gauge fixed C-S theory, at
least in the Coulomb gauge.
Thus, we do not expect that this factor can play an important role in
the computation of the link invariants.
\vskip 1cm
\newsec{ACKNOWLEDGEMENTS}
\vskip 1cm
I wish to thank J. Wess for the warm hospitality at his Institute
and M. Mintchev for fruitful discussions.
\listrefs
\bye